\documentclass{aa} 
	\usepackage{graphicx}
        \usepackage{natbib}
        \usepackage{txfonts}
	
\begin{document}
	
	\title {High Order Correction Terms for The Peak-Peak Correlation Function in Nearly-Gaussian Models}

	\author {A.P.A. Andrade \inst{1,2} \and A.L.B Ribeiro \inst{1} \and C.A. Wuensche \inst{2}}
        \institute {Laborat\'orio de Astrof\'isica Te\'orica e Observacional, Universidade Estadual de Santa Cruz\\
        \email{apaula@uesc.br; albr@uesc.br}
        \and Divis\~ao de Atrof\'isica, Instituto Nacional de Pesquisas Espaciais\\
        \email{alex@das.inpe.br}}

	\date{Received: April 20, 2006;  Accepted: }

	\abstract
	    {One possible way to investigate the nature of the primordial 
	      power spectrum fluctuations is by investigating the statistical properties of the local 
	      maximum in the density fluctuation fields.}
   {In this work we present a study of the mean correlation function, $\xi_r$,
  and the correlation function for high amplitude fluctuations (peak-peak
  correlation) in a slighlty non-Gaussian context.}
   {From the definition of the correlation excess, we compute the Gaussian two-point correlation
	function and, using an expansion in Generalized Hermite polynomials, we estimate the 
	correlation of high density peaks in a non-Gaussian field with generic distribution and 
	power spectrum. We also apply the results obtained to a scale-mixed distribution model, 
	which correspond to a nearly Gaussian model.}
   {The results reveal that, even for a small 
	deviation from Gaussianity, we can expect high density peaks to be much more
	correlated than in a Gaussian field with the same power spectrum. In
	addition, the calculations reveal how the amplitude of the
	peaks in the fluctuations field is related to the existing
	correlations.}
   {Our results may be used as an additional tool to investigate the behavior of the N-point correlation 
function, to understand how non-Gaussian correlations affect the peak-peak statistics 
     and extract more information about the statistics of the density field.}
	

        \maketitle

	\section{Introduction} Investigation of the statistical properties of 
	cosmological density fluctuations is a very useful tool to understanding 
	the origin of the cosmic structure. Roughly, cosmological models to describe
	primordial fluctuations can be divided in two classes: Gaussian and non-Gaussian. 
	The most accepted model for structure formation assumes initial	quantum 
	fluctuations created during inflation and amplified by gravitational effects. 
	The standard inflationary models predicts an uncorrelated random field, with a 
	scale-invariant power spectrum, which follows a nearly-Gaussian distribution 
	\cite{b9,b10}. However, non-Gaussian fluctuations are also allowed in a wide class 
	of alternative models, such as: the multiple interactive fields \cite{b11,b12}, 
	the cosmic defects models \cite{b13} and the hybrid models \cite{b14,b15}. 
	By discriminating between different classes of models, the statistical properties 
	of the fluctuations field can be used to investigate the nature of cosmic structure. 
	However, non-Gaussian models comprehend an infinite range of possible statistics. 
	As a consequence, performing statistical tests of this kind are not a straightforward
	task, since there is no adequate general test for every kind of model. To attack this
	problem, any effort to better understand how the statistical properties of the density 
	fluctuation field affect the observed Universe is welcome, since it may bring extra pieces 
	of information to the investigation of cosmic structure. 
	
	Due to the great importance of characterizing non-Gaussian signatures, many statistical 
	approaches have been used to study the distribution of fluctuations in the cosmic 
	microwave background radiation (CMBR) \cite{b18,b46,b27,b31,b32} and the large scale structure 
	(LSS) \cite{b28,b19,b29,b20}. One possible way to investigate the nature of the primordial 
	power spectrum fluctuations is by investigating the statistical properties of the local 
	maximum in the density fluctuation fields. Since some of the peak properties, such as number, frequency, 
	correlation, height and extrema, are highly dependent upon the statistics of the fluctuation
	field, we can gather information about the statistical distribution function by studying 
	the morphological properties of the fluctuations fields \cite{b8,b22,b35}. Other useful 
	statistical estimator applied to investigate density fluctuations are
	the wavelets tools \cite{b24,b25,b38}, the phase correlations \cite{b17,b26} and the most widely used estimator: 
	the N-point correlations in phase \cite{b23,b6,b21} or density spaces \cite{b5,b30,b33,b34,b37,b39}. 
	
	The extensive use of the two-point correlation function to characterize the statistical 
	properties of the fluctuations field is justified by the mathematical simplicity in the 
	Gaussian condition, since the mean correlation function can be obtained analytically and 
	it completely specifies a Gaussian distribution (as well as its power spectrum). However, 
	this assumption is not true for a non-Gaussian case, where higher order correlations may give a
	significant contribution, despite the great effort demanded to compute a wide range of 
	correlations. However, it is possible to detect primordial non-Gaussianity with a non-zero 
	measure of the N-point correlation $ (n \ensuremath {\ge} 3)$. In order to achieve a good 
	description of non-Gaussian signatures in cosmic structure, many works have been done to
	estimate the three-point correlation function and the related bi or tri-spectrum for a few 
	classes of non-Gaussian models \cite{b6,b32,b27,b21,b41,b42,b43,b44}. It is believed that, 
	with the advent of the CMBR experiments and the high quality surveys in cosmology, the 
	N-point correlation function will be the main statistical descriptor for the cosmic structure. 
	
	In this work we present a technique to extract non-Gaussian components from a two-point 
	correlation function. In presence of a non-Gaussian component, we point out that even 
	between two points it is possible to estimate the influence of higher order correlations 
	for models with different statistical description. By performing the calculation of the 
	high order correction terms for the peak-peak correlation function, we show how the 
	amplitude of peaks is dependent of the correlations involved. We point out that for a 
	non-Gaussian statistics the higher order correlations impose some restrictions to the 
	amplitude of high density peaks.
	
	This paper is divided as follows: in Section 2 we give a general description of a random 
	variable field and estimate the peak-peak correlation function for a Gaussian field. In 
	Section 3 we describe the general treatment to obtain the peak-peak non-Gaussian correlation 
	function. In Section 4 we apply the calculations for a slightly non-Gaussian model in two 
	steps: first we consider an approximated solution for a non-Gaussian model with null high 
	order correlations and finally we estimate the complete solution for a slightly non-Gaussian 
	field. In section 5, we summarise our results and discuss the possibilities to use the 
	peak-peak correlation function as a statistical descriptor of the density fluctuation field.

	\section{Random Variable Fields}
	
	Most of the models for the early universe (i.e. inflation) actually predicts the fluctuation 
	field to be random. This requires that $\delta({\bf x})$ can be treated as a random variable in
	the 3D-space and the assumption that the universe is a random realization from a statistical 
	ensemble of possible universes.

	We define a random variable, \ensuremath{\delta}, using the fact that, instead of knowing its
	exact value, we only know how to measure various values of \ensuremath{\delta_1}, 
	\ensuremath{\delta_2}, ... \ensuremath{\delta_n}, which define a random variable field, 
	under certain experimental conditions. Therefore, a random variable can only be characterized 
	by a certain statistical ensemble of realizations. When we say that a random variable is known, 
	it means that we only know the statistical sample which characterizes it. To completely describe 
	the statistical properties of a random variable, \ensuremath{\delta}, we define the probability 
	density function, $P[\delta]$, which can be obtained from the Fourier transform of the 
	characteristic distribution function, $\Theta_{\delta}$ \cite{b16}:
	
	\begin{equation}\label{eq1}
	 P[\delta]=\frac{1}{2\pi}\int_{-\infty}^{\infty}e^{-iu\delta}\Theta_{\delta}(u)du
	\end{equation}
	
	\noindent The characteristic function, $\Theta_{\delta}(u)$, 
	can be obtained by the McLaurin series for the moments, \cal{m}:
	
	\begin{equation}\label{eq2}
	\Theta_{\delta}(u)=1+\sum_{n=1}^{\infty}\frac{(iu)^n}{n!}m_n, \ \ for: m_n=\left<\delta^n\right>
	\end{equation}

	\noindent  Another possibility to obtain the characteristic function of a random variable
	is to use the  distribution in series of cumulants:
	
	\begin{equation}\label{eq3}
	\Theta_{\delta}(u)=exp\left[\sum_{n=1}^{\infty}\frac{(iu)^n}{n!}k_n\right],  \ for: k_n = \left< \left( \delta -<\delta>\right)^n \right>
	\end{equation}
	
	\noindent  Since the physical importance of the cumulants $k_n$ in Eq. \ref{eq3} 
	decreases as $n$ increases, it is	usual to confine the statistical calculations 
	of random variables to the first few terms of the cumulant distribution series and, 
	for	convenience, set the higher order terms to zero. However, the	calculations presented 
	in the next sections show that, even if the cumulants terms are very small, they can 
	significantly contribute to the statistical description of non-Gaussian fields.
	
	\subsection{Correlations in a Random Field}
	
	The main numerical indication of the correlation degree between random variables
	are the N-order correlation functions. The autocorrelation (or double correlation) 
	for a random variable $\delta$ is defined by:
	
	\begin{equation}\label{eq4}
	K_2[\delta_1\delta_2]=\left<\delta_1\delta_2\right>-\left<\delta_1\right>\left<\delta_2\right>
	\end{equation}
	
	\noindent The triple correlation is similarly defined in terms of all possible
	combinations between the three variables, being:
	
	\begin{eqnarray}\label{eq5}
	K_3[\delta_1\delta_2\delta_3]~~~= & \left<\delta_1\delta_2\delta_3\right>-\left<\delta_1\right>K[\delta_2\delta_3]-\left<\delta_2\right>K[\delta_1\delta_3]\nonumber\\
	& -\left<\delta_3\right>K[\delta_1\delta_2]-\left<\delta1\right>\left<\delta2\right>\left<\delta3\right>
	\end{eqnarray}
	
	Note that, in the case where we have the same three variables (\ensuremath{\delta_1}
	=\ensuremath{\delta_2} = \ensuremath{\delta_3}), the three-point correlation
	function is similar to the third cumulant of the distribution. Higher
	order correlations between several variables can be defined, in a similar
	way, by the difference between all possible correlations involved.
	
	For a statistical process where the correlation functions of order greater
	than one are null, we set the variable described by this correlation function as 
	not random, or deterministic. In the case where the correlation functions of order
	greater than two are null, we have a Gaussian variable. For the	case of correlation 
	functions of order greater than two not completely null, the variable is considered 
	to be non-Gaussian. In this sense, we can say	that a Gaussian random field is a 
	simplified version of a general random field. 
	
	Usually, the cosmological density fluctuation field is statistically described by
	the mean correlation function, \ensuremath{\xi(r)}, applied to a galaxy or a cluster distribution,
	with two-point mean separation defined by $ r$ \cite{b36}. For an
	isotropic and homogenous field, the correlation function is defined as the
	excess of probability for a density field described by a Poisson distribution. Therefore, the probability to find two points, in a volume $dV_1dV_2$, separated
	by a distance $r_{12}$ is given by:
	
	\begin{equation}\label{eq6}
	 dP = n^2dV_1dV_2\left[1 + \xi(r_{12})\right]
	\end{equation}
	
	Describing the fluctuations field in Fourier modes, we have:
	
	\begin{equation}\label{eq7}
	 \xi (r) \equiv  \left< \sum_{k}\sum_{k'}\delta_k \delta_{k'} e^{i(\vec k'-\vec
	 k)\cdot \vec r} e^{-i \vec k \cdot \vec r} \right>,
	\end{equation}
	
	\noindent which is equivalent, in a continuous space, to:
	
	\begin{equation}\label{eq8}
	 \xi (r) = \frac {V}{(2\pi)^3} \int \left|\delta_k\right|^2 e^{-i \vec k \cdot \vec r} d^3
	 \vec k
	\end{equation}
	
	\subsection{High Density Peaks in a Gaussian Random Field}
	
	For a Gaussian random field, the n-dimensional probability density function can 
	be estimated from the Fourier transform of the characteristic function (Eq. \ref{eq1} 
	and \ref{eq2}) defined for moments distribution of s \ensuremath{\leq}2.
	
	\begin{eqnarray}\label{eq9}
	& P^G[\delta_1,\delta_2 ... \delta_n]
	=\frac{1}{(2\pi)^n}   \int_{0}^{\infty}  du_1 \int_{0}^{\infty}du_2
	... \int_{0}^{\infty}du_n \nonumber \\   
	& exp \left[ \displaystyle \sum_{s=1}^{2}\frac{i^s}{s!} \sum_{\alpha,\beta=1}^{n}
	K_s(\delta_\alpha,\delta_\beta)u_{\alpha}u_{\beta}\right]exp{\left[-i\displaystyle
	\sum_{\alpha=1}^{n}
	    u_{\alpha}\delta_{\alpha}\right]}\nonumber\\
	\end{eqnarray}
	
	\noindent The expression above can be reduced to: 
	
	\begin{eqnarray}\label{eq10}
	& P^G[\delta_1,\delta_2 ... \delta_n]~~=~~\frac{1}{(2\pi)^{\frac{n}{2}}} \left[ det
	  \Vert K_2(\delta_\alpha,\delta_\beta) \Vert \right]^{-\frac{1}{2}}\nonumber\\
	& exp \left[-{1\over 2} \displaystyle \sum_{\alpha,\beta=1}^{n} a_{\alpha\beta}
	\left[\delta_{\alpha}-K_1(\delta_{\alpha})\right]
	\left[ \delta_{\beta}-K_1(\delta_{\beta})\right] \right],
	\end{eqnarray}
	
	\noindent where $det\Vert K_2(\delta_\alpha,\delta_\beta) \Vert$ and $\Vert a_{\alpha\beta} 
	\Vert$ are, respectively, the determinant and the inverse of the
	correlation matrix. 
	
	For a bidimensional Gaussian fluctuation field with zero mean, the correlation matrix can 
	be obtained and inverted, resulting in:
	
	\begin{equation}\label{eq11}
	 P^G[\delta_1,\delta_2] = \frac{1}{2\pi\sigma^2}\left(\frac{1}{1-A^2}\right)^\frac{1}{2}e^{-Q[\delta_1,\delta_2]},
	\end{equation}
	
	\noindent where $ A=\frac{\xi_r}{\sigma^2} $, for:\\
	\\
	\noindent $ \xi_r=\left<\delta_1\delta_2\right>$, $ r = |\vec x_2 - \vec
	x_1|$, $\sigma^2=\left< \delta_1^2\right> = \left<\delta_2^2\right>$, and:
	
	\begin{equation}\label{eq12}
	 Q= \frac{1}{2} \left(
	 \frac{1}{1-A^2}\right)\frac{1}{\sigma^2}\left[\delta_1^2+\delta_2^2-2A\delta_1\delta_2 \right],
	\end{equation}
	
	\noindent where $\xi_r$ is equal to $K_2(\delta_1,\delta_2)$, the mean correlation function for a Gaussian field.
	
	To find the correlation function between high density peaks, we calculate the probability of 
	\ensuremath{\delta_1} and \ensuremath{\delta_2} to exhibit density values which are larger than
	the variance field \ensuremath{\sigma}, by a factor \ensuremath{\eta (\eta > 0)}. For a Gaussian 
	field, this probability is given by the integral \cite{b7}:
	
	\begin{equation}\label{eq13}
	 P^G[\delta_1 >\eta\sigma, \delta_2 >\eta\sigma] =
	 \int_{\eta\sigma}^{\infty}d\delta_1\int_{\eta\sigma}^{\infty}d\delta_2P^{G}[\delta_1,\delta_2].
	\end{equation}
	
	The integral above can be obtained by substituting Eq. \ref{eq9} into Eq. \ref{eq13}. For a 
	weakly correlated field, where $A= \frac{\xi_r}{\sigma^2} << 1$, and high density peaks, 
	$\eta >1$, the integral above will be:
	
	\begin{eqnarray}\label{eq14}
        & P^G[\delta_1 >\eta\sigma, \delta_2 >\eta\sigma]=\nonumber\\
	& \int_{\eta\sigma}^{\infty}d\delta_1\int_{\eta\sigma}^{\infty}d\delta_2\left[\frac{1}{\sigma^2}
	\displaystyle \sum_{p=0}^{1}F^{p+1}\left(\frac{\delta_1}{\sigma}\right)F^{p+1}\left(\frac{\delta_2}{\sigma}\right)
	\right]\frac{(2A)^p}{p!}, \nonumber\\
	\end{eqnarray}
	
	\noindent where F is the Err function (erf):
	\begin{equation}\label{eq15}
	F^{p+1}(Z)= \frac{(-i)^p}{2\pi}
	\int_{-\infty}^{\infty}\lambda^p e^{\left[ -i \lambda z -\frac{z^2}{2} \right]} d\lambda.
	\end{equation}
	
	\noindent The final result is:
	\begin{equation}\label{eq16}
	 P^G[\delta_1 >\eta\sigma, \delta_2 >\eta\sigma] =
	\left[\frac{1}{\sqrt{2\pi}} \frac{1}{\eta} e^{-\frac{\eta^2}{2}}\right]^2 (1 +2A\eta^2).  
	\end{equation}
	
	Redefining expression \ref{eq16}, the probability to find peaks $\delta_1$ and $\delta_2$ 
	with density $\eta$ times the variance, $\sigma$, is:
	
	\begin{equation}\label{eq17}
	 P^G[\delta_1 >\eta\sigma, \delta_2 >\eta\sigma] \equiv
	[P^G(\delta>\eta\sigma)]^2 [1 + \xi_{\eta}^G(r)], \\ 
	\end{equation}
	
	\noindent where $[P^G[\delta>\eta\sigma)]]^2$ is the mean probability to find high density peaks in a bi-dimensional random field, and $[1 + \xi_{\eta}^G(r)]$ is the probability excess expressed in terms of the mean correlation function, $\xi_r$. Then, for weakly correlated fields: 
	
	\begin{equation}\label{eq18}
	\xi_\eta^G(r) \equiv 2A\eta^2 \approx e^{2A\eta^2}-1 = exp\left[2\left(\frac{\eta}{\sigma}
	\right)^2\xi_r\right]-1.
	\end{equation}

	\section{Correlations in a Generic Non-Gaussian Field}
	
	The correlation function of high density peaks in a non-Gaussian field can also be computed 
	using	Eq. \ref{eq13}, except that we have to consider the non-Gaussian probability of finding 
	both peaks $\delta_1$ and $\delta_2$, the $P^{NG}[\delta_1,\delta_2]$. This probability can be 
	estimated using Eq. \ref{eq9}, performing the sum over terms of order higher than 2. Since the 
	summation is also computed for additional terms, the result can be synthesized by:
	
	\begin{eqnarray}\label{eq19}
	P^{NG}[\delta_1 >\eta\sigma, \delta_2 >\eta\sigma]~~~~~= &
	 [P^G(\delta>\eta\sigma)]^2 \nonumber\\
	& \left[1 + \xi_{\eta}^G(r)+ C_{\eta}^{NG}\right],
	\end{eqnarray}
	
  \noindent which results in:

	\begin{eqnarray}\label{eq20} 
	P^{NG}[\delta_1 >\eta\sigma, \delta_2 >\eta\sigma] & = \nonumber\\
	\left[\frac{1}{\sqrt{2\pi}} \frac{1}{\eta} e^{-\frac{\eta^2}{2}}\right]^2 & (1 +2A\eta^2+C_{\eta}^{NG}),
	\end{eqnarray}
	
	\noindent where the factor $C_{\eta}^{NG}$ carries the corrections terms of higher order (s \ensuremath{\ge 3}).

	One intriguing question we could ask is how important the high order residual terms for 
	a slightly non-Gaussian statistics are . First, we could consider the approximated case 
	in which the extra calculation in Eq. \ref{eq9} were avoided, so correlation terms of order (s \ensuremath{> 2})
        could be neglected ($C_{\eta}^{NG} \sim 0 $) and the non-Gaussian 
	peak-peak correlation function would be reduced to:
	
	\begin{equation}\label{eq21}
	\xi_\eta^{NG}(r) \approx exp \left[2\left(\frac{\eta}{\sigma}\right)^2\xi_r^{NG}\right]-1,
	\end{equation}
	
	This solution is similar to the calculation for a Gaussian random field, except for the fact 
	that we were considering a non-Gaussian mean correlation function. 
	
	At this point we want to assess the robustness of the assumption $C_{\eta}^{NG} \sim 0 $, 
	stated in the previous paragraph. For this purpose, we compute the higher order correction terms 
	considering a slightly non-Gaussian component, this means a very small contribution to 
	correlations of order greater than two. However, the non-Gaussian probability described in 
        Eq. \ref{eq9} can not be calculated 
	using the Fourier transform of the characteristic function, since there is no analytical 
	solution for that expression. One possible way to obtain $P^{NG}[\delta_1,\delta_2]$, as 
	suggested by \cite{b16}, is to expand it in a series of Generalized Hermite polynomials, H: 
	
	\begin{eqnarray}\label{eq22}
	& P^{NG}[\delta_1,\delta_2 ...\delta_n]= P^G[\delta_1,\delta_2...\delta_n] \nonumber\\
	& \left[ 1 + \displaystyle \sum_{s=3}^{\infty} \frac{1}{s!}
	\displaystyle \sum_{\alpha,\beta...\omega =1}^{n}b_s(\delta_\alpha,\delta_\beta...\delta_\omega)H_{\alpha,\beta...\omega}\left[\delta -\kappa_1 (\delta) \right]\right], \nonumber\\
	\end{eqnarray}
	
	\noindent where $b_s$ is the quasi-moment function and \ensuremath{\kappa_1} is the first 
	cumulant of the distribution. The definitions of $b_s$ and $H$ are given in Appendix A.
	
	By Eq. \ref{eq22}, the non-Gaussian probability is expressed in terms of a Gaussian 
	probability added to higher order $(s \ge 3)$ correction terms, which are related to the 
	deviation from Gaussianity. Combining equations \ref{eq13}, \ref{eq19} and \ref{eq22},we can estimate 
	the higher order corrections term for a bi-dimensional using the following:

	\begin{eqnarray}\label{eq23}
	& C^{NG}_{\eta}~~=~~\int_{\eta\sigma}^{\infty} \int_{\eta\sigma}^{\infty} P^G[\delta_1,\delta_2] d\delta_1 d\delta_2 \nonumber \\
        & \left[ \displaystyle \sum_{s=3}^{\infty} \frac{1}{s!} \displaystyle
        \sum_{\alpha,\beta...\omega
        =1}^{2}b_s(\delta_\alpha,\delta_\beta)H_{\alpha,\beta}\left[\delta
        -\kappa_1 (\delta) \right] \right].\nonumber\\
	\end{eqnarray}

Note that the calculation of $C^{NG}_\eta$ starts at s \ensuremath{\ge 3}, ensuring that the terms used in the expansion are related to the non-Gaussian contributions. 
Details of the calculation involved in this higher order terms are also presented in Appendix A. 

	\section{Correlations in a Nearly-Gaussian Field}

        \subsection{The Mixture Model}

	In order to estimate how the high order terms affect the correlation between high density 
	peaks, we estimated the	$\xi_{\eta}$ for a Gaussian and a slightly non-Gaussian field, computing
	the approximated solutions (Eq. \ref{eq21}) and the full calculation of the expansion (Eq. \ref{eq22})
	until $s \leq 6$ order. For this comparison, we have considered a mixed probability distribution,
	as proposed by Ribeiro, Wuensche \& Letelier \cite {b3}, hereafter (RWL).

	The general procedure to create a wide class of non-Gaussian models is to admit the
	existence of an operator which transforms Gaussianity into non-Gaussianity according to 
	a specific rule. An alternative approach to study non-Gaussian fields
	was proposed by 
	RWL, in which the PDF is treated as a mixture: $P(\psi)= (1- \alpha)
	f_1(\phi) + \alpha f_2(\phi)$, 
	where	$f_1(\phi)$ is a (dominant) Gaussian PDF and $f_2(\phi)$ is a	second distribution, 
	with $0 \le \alpha \le 1$. The $\alpha$ parameter gives the absolute level of Gaussian deviation,
	while $f_2(\phi)$ modulates the shape of the resultant non-Gaussian distribution. RWL used this 
	mixed scenario to probe the evolution of galaxy	cluster abundance in the universe and found 
	that even at a small level of non-Gaussianity $(\alpha \sim 10^{-4}-10^{-3})$ the mixed 
	field can introduce significant changes in the cluster abundance rate.
	
	The effects of such mixed models in the CMBR power spectrum, combining a Gaussian adiabatic
	 field with a second, non-Gaussian isocurvature fluctuation field, to produce a
	positive skewness density field was discussed by Andrade, Wuensche \&
	 Ribeiro \cite{b1} (hereafter AWR04). In this
	approach, they adopted a scale dependent mixture parameter and a power-law
	initial spectrum to simulate the CMBR temperature and polarization power spectra for a 
	flat \ensuremath{\Lambda}-CDM model, generating a large grid of cosmological parameters
	combination. The choice of a scale-dependent mixture is not	unjustified, since it could 
	fit both CMB and high-$z$ galaxy clustering in the Universe (e.g. AWR04; Avelino \& Liddle 2004; 
	Mathis, Diego \& Silk 2004). At the same time, in mixed scenario, the scale-dependence acts in
	order to keep a continuous mixed field inside the Hubble horizon. Simulation results show how the shape and amplitude of the fluctuations in CMBR depend 
	upon such mixed fields and how a standard adiabatic Gaussian field can	be distinguished 
	from a mixed non-Gaussian one. They also allow one to quantify the contribution of the 
	second component. By applying a $\chi^2$ test on recent CMBR observations, the contribution 
	of the isocurvature field was estimated by Andrade, Wuensche \&
	 Ribeiro \cite {b2} (hereafter AWR05) as \ensuremath{\alpha}$_{0}$= 0.0004 \ensuremath{\pm} 
	0.00030 with 68\% confidence limit. In the present work, we also investigate 
	the predictions of scale-dependent mixed non-Gaussian cosmological density fields for the
	peak-peak correlation function.

        \subsection{The Two-Point Correlation Function}

	To obtain the mean correlation function, $\xi_r$, we have computed the Fourier transform 
	of the power spectrum related to a pure Gaussian field and a mixed non-Gaussian PDF. 
	In this sense, we rewrite Eq. \ref{eq8}, which is equivalent to:
	 
	\begin{equation}\label{eq24}
	 \xi (r) = \frac {V}{(2\pi)^3} \int_{0}^{\infty} P(k)k^2dk \int_{0}^{2\pi}d\phi\int_{-1}^{1}e^{-ikru}du,
	\end{equation}
	
	\noindent being $u=cos \theta$. For an isotropic field, we have:
	
	\begin{equation}\label{eq25}
	 \xi (r) = \frac {V}{(2\pi)^3} \int_{0}^{\infty} P^{NG}(k)\frac{sin(kr)}{kr}k^2dk
	\end{equation}
	
	For the mixture correlation function, we consider a mixed-scale power spectrum described as:
	
	\begin{equation}\label{eq26}
	P^{Mix}(k)=A_{n}P(k)[1+M(\alpha_0)k] = \beta k^n[1+M(\alpha_0)k]
	\end{equation}
	
	This power spectrum was estimated by the correlated mixed-model which considers a possible 
	mixture, inside the horizon, between fluctuations of an adiabatic Gaussian field and an 
	isocurvature non-Gaussian one (AWR04). In this model, $\alpha_0$ is the mixture parameter 
	that modulates the contribution of the isocurvature field (\ensuremath{\alpha}$_{0} = 
	(0.00042 \ensuremath{\pm} 0.00030)$) as estimated from recent CMB observations. 
	For a null $\alpha_0$, the field is purely Gaussian with a simple power law	spectrum. $A_{n}$ 
	is the normalization constant of the primordial	power spectrum, estimated as $\sim 1.3 
	\times 10^{-11}$ for recent CMBR observations (AWR05). $M(\alpha_0)$ is the mixed term, 
	which accounts for the statistical effects of the second component in the power spectrum, 
	expressed as a functional of both distributions, $f_{\mathit{1}}$ and $f_{\mathit{2}}$:
	
	\begin{equation}\label{eq27}
	M^{mix}(\alpha_0,k)\equiv\int_\nu[(1-\alpha_0k)f_1(\nu)+\alpha_0kf_2(\nu)]\nu^2\; d\nu
	\end{equation}
	
	Evaluating the integral in Eq. \ref{eq25} for a mixed power spectrum, we find an expression 
	for the mean correlation function:
	
	\begin{equation}\label{eq28}
	\xi^{Mix}(R)=A_{n}\left[\frac{1.76\times 10^{-2}}{3\pi
	\left(\frac{R}{R_0}\right)} + \frac{5.76\times 10^{-4}M(\alpha_0)}{3\pi \left(\frac{R}{R_0}\right)^2} \right],
	\end{equation}
	
	\noindent where $R_0$ is about 25 Mpc, the mean correlation width for galaxy clusters. 
	
	In Figure 1, we show the mean correlation function estimated for a pure Gaussian field, 
	$ M(\alpha_0) =0$, for a mixed PDF, $M(\alpha_0) = \alpha_0 \left( \frac{e^2}{2}-1\right)$, 
	and also the individual contribution of the non-Gaussian field for $\alpha_0 \sim 10^{-3}$. 
	In this plot, it is possible to observe the importance of even a small contribution of a
	the second component to the mean correlation function. For $ \frac{R}{R_0} \leq 1 \times 10^{-3}$ 
	the non-Gaussian component dominates the correlation function. This	behaviour illustrates 
	the excess of power in small scales, as observed in the CMB angular power spectrum in	the mixed 
	context (AWR04).

 	\begin{figure}\label{fig1}
 	\begin{center}
 	\includegraphics[width=3.4in, height=2.6in]{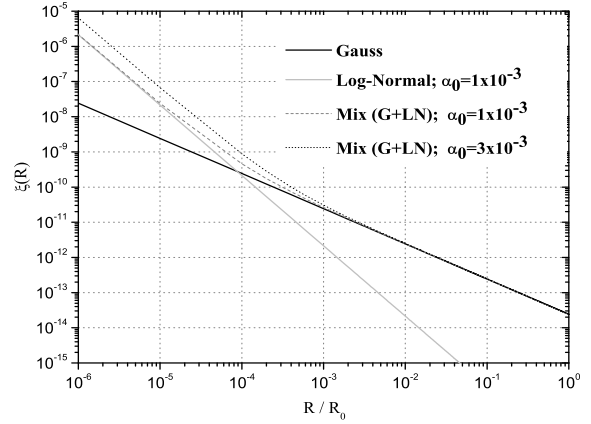}
 	\caption {\small The mean two-point correlation function estimated for
non-correlated or weakly correlated fields: pure Gaussian, pure log-normal and Mixed field with $\alpha_0 \sim 10^{-3}$.}
 	\end{center}
 	\end{figure}
		
	Inserting the mean correlation function described in Eq. \ref{eq28}
	into the approximated expression of the correlation function for high
	density peaks (Eq. \ref{eq21}) for a few classes of PDFs, we estimate
	the functions $\xi_\eta$ plotted in Figure 2. In this plot,
	we show that the effect of the second component is still observed, and
	that the correlation function for high density peaks is also sensible
	for different non-Gaussian distributions. Comparing Figures 1 and 2,
	we see how the high density peaks can be much more correlated than the
	mean field for a non-Gaussian case, especially in small scales,
	$\frac{R}{R_0} \sim 10^{-5}$ where $\xi_\eta^{Mix}$ is nearly two 
	orders of magnitude greater than the mean correlation.
	
 	\begin{figure}\label{fig2}
 	\begin{center}
 	\includegraphics[width=3.4in, height=2.6in]{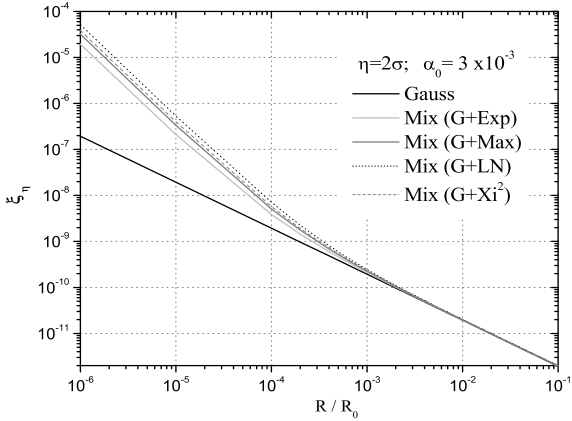}
 	\caption {\small The two-point correlation function computed for $2\sigma$ density peaks 
 	in both a pure Gaussian and mixed context ($\alpha_0 \sim 10^{-3}$). For this estimation, 
 	we used: $ M^{Gaus}(\alpha_0) =0$; $ M^{Exp}(\alpha_0) =\alpha_0$; $ M^{Max}(\alpha_0) 
 	=\alpha_0\left( 3 \sqrt{\frac{\pi}{2}}-1\right)$; $ M^{Xi^2}(\alpha_0) =2\alpha_0$ and  
 	$M^{LN}(\alpha_0) = \alpha_0 \left( \frac{e^2}{2}-1\right)$.}
 	\end{center}
 	\end{figure}
	
	With the help of a program that performs algebraic and numerical calculations, we actually computed
	$C_{\eta}^{NG}$, as indicated in Eq. \ref{eq23}, for correlations 
	up to the 6-th order ($ s \le 6$). This	limit was set in order to keep a meaningful 
	non-Gaussian distortion, avoiding more time consuming calculations. We
	do not present in this section the full computed expression since it contains hundreds of non-linear terms in the 
mean correlation function $(\xi_r)$ and	$\eta$, where the coefficients are
the high order correlations (or cumulants). In fact, the $C^{NG}_{\eta}$ is more accurately described as $C^{NG}_{\eta}(\xi_r)$. In Appendix A, we show the steps to calculate the quasi-moment function, $b_{lm}$, and
	the Hermite Polynomials, $H_{lm}$. 

        In general, correlations of very high order tend to zero \cite{b16}, the most extreme case being the normal distribution, where all cumulants of $s \ge 3$ are null. Deviations from Gaussianity are set by the increment of non-vanishing cumulants in the expansion of $C_{\eta}^{NG}$. A possible question that may be raised is related to the convergence and normalization of the expansion in Eq. \ref{eq23} in which some terms are set as non-vanishing. However, investigation 
	of general nearly gaussian deviations have already been performed by the use of the Edgeworth 
	expansion in one-dimension by setting the function to zero in higher orders terms
	and normalizing it appropriately \cite {b48}. Following these authors, we have considered, as a work
	hypothesis, the following behavior for such coefficients: we have set the terms $s \ge 7$ 
	to zero and the cumulants involved in the quasi moment function to $10^{-3}$ for $3 \le s \le 6$.

        A very interesting result obtained is summarized in Figure
	3. For this plot, in order to follow the absolute behavior of the corrections
	terms in the nearly Gaussian field, we have set the normalization value of the mean
	correlation function to one, $(\xi_r \sim 1)$ and explore the $\eta$
	dependence of $C^{NG}_{\eta}(\xi_r)$ with the purpose to understand
	how the amplitude of the fluctuations is related with the high order correlations. 
	While computing only third-order terms, the $C_{\eta}^3$ describes a gradual enhancement
	in the correlation function. For fourth and sixth-order terms, the 
	$C_{\eta}^4$ and $C_{\eta}^5$, in the case of small Gaussian deviation, what is observed is that the $C_{\eta}^4$ and $C_{\eta}^5$ just overlap each other, meaning that the fifth order correlations do not contribute significantly. 
	However, the fourth order correlation describes peaks of maximum density allowed by a weakly
	correlated field at about $\eta \sim 1.8~\sigma$. For higher order correlations, the
	$C_{\eta}^6$ shows a very large increment in the two point correlation for densities about 
$2.5~\sigma$, and a nearly null contribution to peaks with amplitudes ($\eta \ge 3\sigma$). 

	In this sense, we conclude that one should not expect very high density peaks for the specified, 
	slightly non-Gaussian, correlated field. However, factors or order $3 \le s \le 6$, are quite significant for peaks with amplitude up to $2.5~\sigma$ and are too far from a null correction, even if we consider weakly correlated fields.
        Then, we can not consider the simplified solution in Eq. \ref{eq21} as a good
	description of the two-point correlation function for high density
	peaks in despite of our choice of considering just a small deviation from Gaussianity.
        It is important to note that this result is independent of the power spectrum or the mean
	correlation function. It only shows the influence of higher order
	correlations in the amplitude of the field of fluctuations.

	Analyzing the relation between $C_{\eta}$ and $\eta$ we gain 
some insight about the amplitude of such high order terms, since
$C_{\eta}$ controls the amplitude of the permitted peaks in the fluctuations
field. Increasing values of correlations with $s > 6 $ imply a high
probability of very high amplitude (very rare) peaks, which contradict the observations of large scale structures. However,
when we impose correlation levels of the order $10^{-3}$ up to sixth order, we
  favour the existence of peaks up to $3\sigma$, what is very reasonable for a
  nearly Gaussian field.

	In Figure 4 we show the behaviour of the two-point correlation
	function estimated for a Gaussian, $\xi_{\eta}^{Gauss}$; for a mixed approximated
	solution, $\xi_{\eta}^{Mix}$; and for the non-Gaussian complete
	solution, $\xi_{\eta}^{Mix} + C_{\eta}^{NG}(\xi_r)$ . In this plot, we have
	set the amplitude threshold of $\eta \sim 2$ and test the dependence
	on $\xi_r$. The observed effect of $C^{NG}_{\eta}(\xi_r)$ is to
	amplify the correlations between high density peaks. This result is
	valid for the case of an increasing mean correlation function, and has
	no dependence of the mixed model. While non-Gaussian deviation tends
	to add non-vanish high order correlations, we conclude that we can
	expect high density peaks to be much more correlated even in a
	slightly non-Gaussian model. 

 	\begin{figure}\label{fig3}
 	\begin{center}
 	\includegraphics[width=3.4in, height=2.6in]{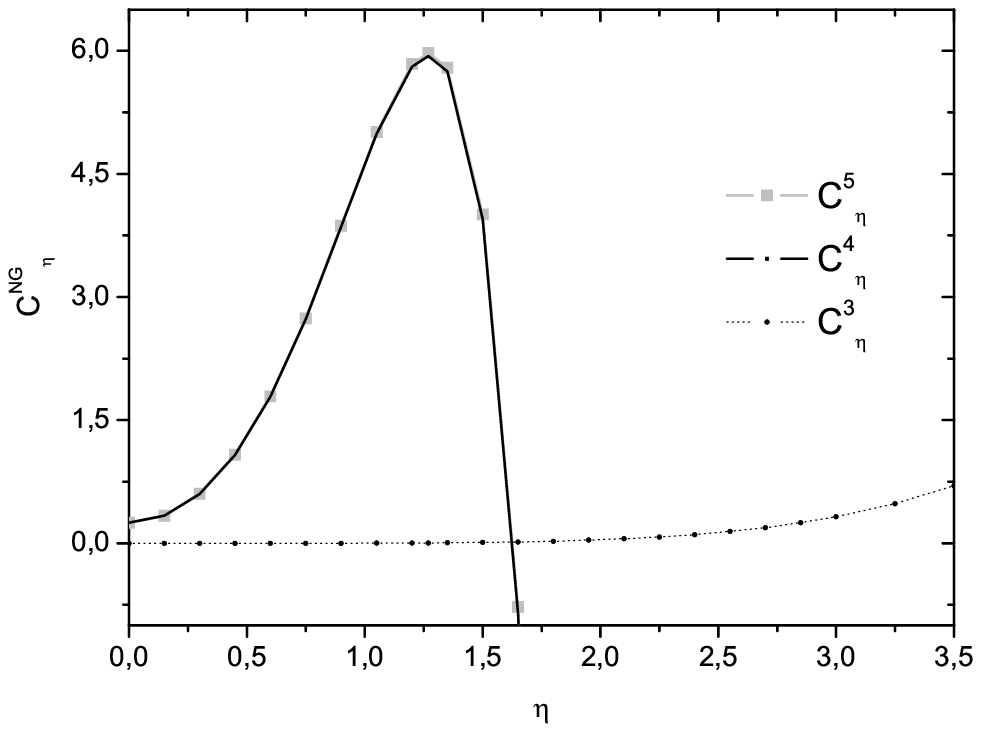}
 	\includegraphics[width=3.4in, height=2.6in]{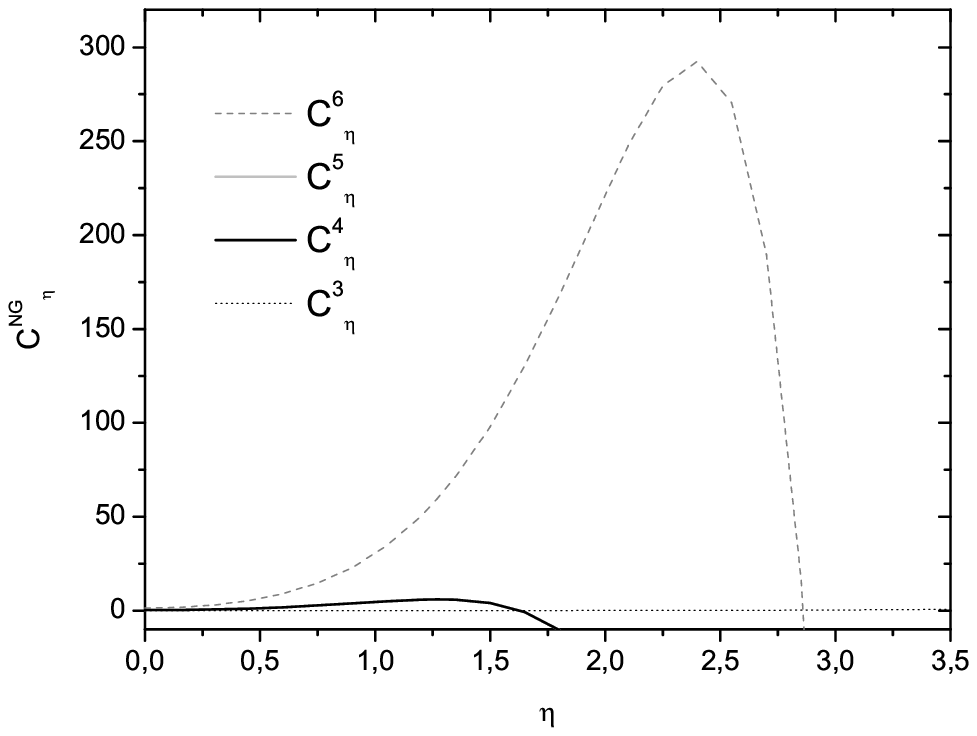}
 	\caption {\small The behaviour of the higher order factor, $C_{\eta}^{NG}$, for 
 	a fixed mean correlation ($\xi_r \sim 1$). The curves in A) show the factor 
 	estimated for third order correlation, $C_{\eta}^{3}$, fourth, $C_{\eta}^{4}$, 
 	and fifth, $C_{\eta}^{5}$. In B) the sixth order factor, $C_{\eta}^{6}$, is also plotted.}
 	\end{center}
 	\end{figure}
	
 	\begin{figure}\label{fig4}
 	\begin{center}
 	\includegraphics[width=3.4in, height=2.6in]{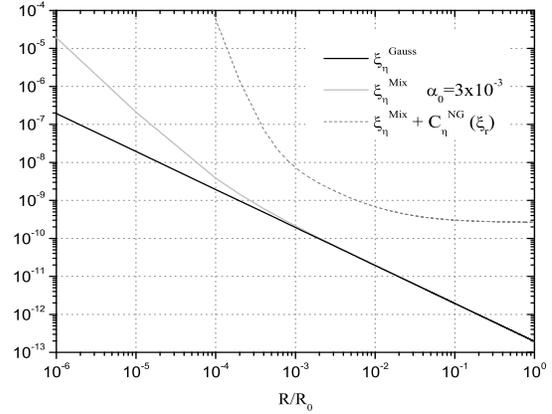}
 	\caption {\small The behaviour of the two-point correlation
	function is shown for three estimated cases: a pure Gaussian, $\xi_{\eta}^{Gauss}$ (lower curve, orange),
	a mixed approximated solution, $\xi_{\eta}^{Mix}$ with $\alpha_0 \sim 3 \times 10^{-3}$ (mid curve, black); 
	and for the complete non-Gaussian solution, $\xi_{\eta}^{Mix} + C_{\eta}^{NG}(\xi_r)$, estimated for a PDF 
	of the type: (Gauss + Exp) with $\alpha_0 \sim 3 \times 10^{-3}$}
 	\end{center}
 	\end{figure}
	
	\section{Discussion}
	
	In this work, we have estimated the two-point mean correlation function 
	and the peak-peak correlation function for the density field.
	In the Gaussian case, the calculations are simplified,
	since the Fourier transform of the power spectrum completely describes
	the random variable. However, for a non-Gaussian field, the
	calculations are much more complicated, since high order correlations
	between these two-points may not vanish and strongly contribute to the final
	function, even in a small deviation from the Gaussian case.   

	In this work, we showed that, when considering a mixed model, both the mean correlation and 
	the peak-peak correlation functions are much more intense in small scales than in the Gaussian case. This result can be
 	particularly relevant since it is generally accepted that galaxies form in high density regions. 
	In addition, we conclude that the peak-peak correlation function is quite sensitive to the PDF of the 
	fluctuations field, especially for a mixed model. This result suggests that it is possible to use the 
	peak-peak correlation function as a test for the nature of cosmic structures. 
        Nevertheless, we have to be careful when approximating terms for high order correlations, since the peak-peak correlation function is very sensitive to the correction terms. We also point out that correlations of order \ensuremath{> 2} can be a very important tool to characterize non-Gaussian fields and definitely deserve a deeper 
	investigation. 	Estimating high order correlations allows us to investigate the behaviour of the N-point correlation 
	function, as well as gather more information about the amplitude of the expected high density fluctuation 
	field. As seen in Figure 3, correlations may restrict the amplitude of the density peaks. Furthermore, from 
	the amplitude of the peaks found in the density field (in CMB or LSS fluctuations) we are able to extract 
	more information about the statistics of the density field. It is also
	good to remember that the presence of correlations of order $ s \ge 3$ lead to formation of structures in earlier times
than would be expected for a model with the same power spectrum but with weaker spatial correlations. The resuls
presented in this paper may be used to set new constraints in structure formation models.

One possible application of this method on investigations of primordial 
 non-Gaussianity could be implemented in the search for maximum amplitude
 fluctuations of the full sky CMB temperature maps, such those derived from WMAP \cite {b49}
 and the future Planck mission \cite {b50}. However, variations in the
 number of density peaks and their correlation may as well be related to 
 non-Gaussian Galactic foregrounds or other contaminants. Once detected such non-Gaussian trace,
we have to be very careful before assigning it to a primordial origin. To avoid misunderstanding,
the investigation of the peaks statistics in CMB datasets should be realized over 
several datasets and different frequencies. Indeed, one possible manner to 
minimize the foregrounds effects is to analyse the most sensitive and 
cleaned map, such the WMAP three-year di-biased internal linear combination,
 WMAP-DILC map, or the WMAP coadded map, with the combination of Q+V+W frequencies.
 This analysis will be the next test applied to the non-Gaussian mixed model.

\begin{acknowledgement}

	APAA thanks the financial support of FAPESB and CNPq, 
	under grant 1431030005400. ALBR thanks 
	the financial support of CNPq, under grants 470185/2003-1 and 306843/2004-8. 
	CAW was partially supported by CNPq grant 307433/2004-8.

\end{acknowledgement}

\appendix
\section{Calculation of $b_s$ and $H$}

To obtain the non-Gaussian probability for a multi-dimensional case we have to
perform the calculation in Eq. \ref{eq22} expanding the Hermite Polynomials and
the quasi-moment function. One possible way to obtain the quasi-moment functions, 
$b_s$, is relating to the correlation function, since: 
	
	\begin{eqnarray}\label{eqA1}
	exp\left[\ \displaystyle \sum_{s=3}^{\infty} \sum_{\alpha,\beta,...,\omega=1}^{n}
	  K_s(x_{\alpha},x_{\beta},..., x_{\omega})z_{\alpha}z_{\beta}...z_{\omega} \right]\nonumber\\
	= 1+ \displaystyle \sum_{s=3}^{\infty}\frac{i^s}{s!}\sum_{\alpha,\beta,...,\omega=1}^{n}b_s(x_{\alpha},x_{\beta},...,
	  x_{\omega})z_{\alpha}z_{\beta}...z_{\omega}.
	\end{eqnarray}
	
\noindent The Generalized Hermite polynomials can be obtained by the
definition: 
	\begin{eqnarray}\label{eqA2}
	H_{\alpha\beta...\omega}[x]= e^{\phi[x]}\left(
	-\frac{\partial}{\partial x_{\alpha}}\right) \left(
	-\frac{\partial}{\partial x_{\beta}}\right)...\left(
	-\frac{\partial}{\partial x_{\omega}}\right)e^{-\phi[x]},\nonumber\\
	\end{eqnarray}
	
	\noindent where: $\phi[x]=\frac{1}{2}\sum_{\alpha,\beta=1}^{n}a_{\alpha\beta}x_{\alpha}x_{\beta}$, 
	being $||a_{\alpha\beta}||$ the elements of the inverse correlation matrix $||k_2(t_{\alpha},
	t_{\beta})||$. The unidimensional case of Eq. \ref{eq24} is known as Edgeworth series, while the 
	bidimensional case corresponds to:
	
	\begin{eqnarray}\label{eqA3}
	& P^{NG}[\delta_1,\delta_2]=\nonumber\\
	& \left[1+ \displaystyle \sum_{s=3}^{\infty}\frac{1}{s!}\sum_{l+m=s}^{}b_{lm}H_{lm}\left[\delta-\kappa_1\left(\delta\right)\right]\right]P^{G}\left[\delta_1,\delta_2\right],\nonumber\\
	\end{eqnarray}
	
	\noindent where: $b_{lm}= b_{l+m}[\underbrace {\delta_1 \cdots
	\delta_1}_{l \ \textrm{times} }; \underbrace {\delta_2 \cdots \delta_2}_{m \
	\textrm {times}}]$, \\
      
        \noindent and:  $H_{lm}=H_{\underbrace{1 \cdots 1}_{ l \
        \textrm{times}};\underbrace{2 \cdots 2}_{ m \ \textrm{times}}}$.\\ \\

Performing the calculations for the quasi-moment function we have:\\

\noindent $b_3=K[\delta_1,\delta_2,\delta_3]; $ \\
\noindent $b_4=K[\delta_1,\delta_2,\delta_3,\delta_4];$ \\
\noindent $b_5=K[\delta_1,\delta_2,\delta_3,\delta_4,\delta_5];$ \\
\noindent $b_6=K[\delta_1,\delta_2,\delta_3,\delta_4,\delta_5,
  \delta_6]+10\left[K_3[\delta_1,\delta_2,\delta_3]K_4[\delta_4,\delta_5,\delta_6]\right].$
  \\ 

\noindent Note that the first five terms of $b_s$ are equivalent to the correlation function, 
only the sixth term have additional terms. As an example, for $s = 3$, we obtain four different 
forms of $b_3$:\\

\noindent $ b_{30}=K_3[\delta_1,\delta_1,\delta_1,] = \left < \delta_1^3 \right>;$ \\
\noindent $ b_{03}=K_3[\delta_2,\delta_2,\delta_2,] = \left < \delta_2^3 \right>;$ \\
\noindent $ b_{21}=K_3[\delta_1,\delta_1,\delta_2,] = \left < \delta_1^2\delta_2 \right>;$\\
\noindent $ b_{12}=K_3[\delta_1,\delta_2,\delta_2,] = \left < \delta_1\delta_2^2 \right>$.\\

Proceeding in a similar manner, we find five terms for $b_4$, six for $b_5$
and seven terms for $b_6$. 

The best way to find the Hermite Polynomials is using expression \cite{b16}:

\begin{equation}
y_\alpha=y_\alpha[x]=\frac{\partial \phi[x]}{\partial
  x_\alpha}=\sum_{\beta=1}^{n}a_{\alpha\beta}x_\beta, \ \ \ \ being:
\end{equation}

\noindent $H_\alpha ~~~~~~~~= y_\alpha, $  \\
$ H_{\alpha\beta} ~~~~~~= y_\alpha y_\beta - a_{\alpha\beta}, $  \\
$ H_{\alpha\beta\gamma} ~~~~~= y_\alpha y_\beta y_\gamma - a_{\alpha\beta} y_\gamma -
a_{\alpha\gamma} y_\beta - a_{\gamma\beta} y_\alpha  $  \\ 
$~~~~~~~~~~~~~= y_\alpha y_\beta y_\gamma - \left\{ a_{\alpha\beta} y_\gamma
\right\}_3,\nonumber $  \\
$ H_{\alpha\beta\gamma\delta} ~~~~= y_\alpha y_\beta y_\gamma y_\delta
-\left\{a_{\alpha\beta}y_\gamma y_\delta \right\}_6 - \left\{ a_{\alpha\beta}
a_{\gamma\delta} \right\}_3, $  \\
$ H_{\alpha\beta\gamma\delta\omega} ~~= y_\alpha y_\beta y_\gamma y_\delta
y_\omega - \left\{ a_{\alpha\beta}y_\gamma y_\delta y_\omega \right\}_{10} - \left\{ a_{\alpha\beta} a_{\gamma\delta} y_\omega \right\}_{15}, $ 
$ H_{\alpha\beta\gamma\delta\omega\theta} ~= y_\alpha y_\beta y_\gamma y_\delta
y_\omega y_\theta - \left\{ a_{\alpha\beta} y_\gamma y_\delta y_\omega
y_\theta \right\}_{15} - $ \\
$ ~~~~~~~~~~~~~~~~~~~~~~~~~~~~~~\left\{ a_{\alpha\beta} a_{\gamma\delta} y_\omega
y_\theta   \right\}_{45} - \left\{ a_{\alpha\beta} a_{\gamma\delta}
a_{\omega\theta} \right\}_{15}.$ \\ 

\noindent where we have used the notation $\left\{ i \right\}_s$ to indicate a simetrization set.

To perform the calculation above for $b_{lm}$ and $H_{lm}$ up to $(l+m = 6)$
and perform the integration in $d\delta_1$ and $d\delta_2$, we used a software for 
algebraic and numerical calculations. 

\end{document}